\documentclass[runningheads]{llncs}

\usepackage[T1]{fontenc}

\usepackage{latexsym}
\usepackage{amsfonts}
\usepackage{graphicx}
\usepackage{hyperref}
\usepackage{color}

\def\Z{{\mathbf Z}}

\begin{document}

\title{Tropical cryptography III: digital signatures}

\author{Jiale Chen \inst{1}  \and Dima Grigoriev \inst{2}  \and Vladimir Shpilrain \inst{3}}

\authorrunning{Chen, Grigoriev, Shpilrain}

\institute{Department of Mathematics, The City College of New York, New York,
NY 10031 \email{jchen056@citymail.cuny.edu} \and CNRS, Math\'ematiques, Universit\'e de Lille, 59655, Villeneuve d'Ascq, France \email{Dmitry.Grigoryev@univ-lille.fr} \and Department of Mathematics, The City College of New York, New York, NY 10031 \email{shpilrain@yahoo.com}}

\maketitle

\begin{abstract}
We use tropical algebras as platforms for a very efficient digital signature protocol. Security  relies on computational hardness of factoring one-variable tropical polynomials; this problem is known to be NP-hard. We also offer  countermeasures against recent attacks by Panny and by Brown and Monico.


\end{abstract}


\section{Introduction}

In \cite{tropical}, \cite{tropical2}, we employed {\it tropical
algebras} as platforms for cryptographic schemes by mimicking
some well-known classical schemes, as well as newer schemes like \cite{semidirect1}, \cite{semidirect2},  in the ``tropical" setting.
What it means is that we replaced the usual operations of addition
and multiplication by the operations $\min(x,y)$ and $x+y$,
respectively.

An obvious advantage of using tropical algebras as
platforms  is unparalleled efficiency because in tropical schemes,
one does not have to perform any multiplications of numbers since
tropical multiplication is the usual addition, see  Section
\ref{Preliminaries}. On the other hand, ``tropical powers" of an
element may exhibit some patterns, even if such an element is a matrix
over a tropical algebra. This weakness was exploited in \cite{Kotov}
to arrange a fairly successful attack on one of the schemes in
\cite{tropical}.

In this paper, we offer a digital signature scheme that uses tropical algebra of
one-variable polynomials. Security of the public key in this scheme is based on computational hardness of factoring one-variable tropical polynomials. This problem is known to be NP-hard, see \cite{Kim}.

Since the first version \cite{original} of our paper was put online in September 2023, Panny \cite{Panny} and Brown and Monico \cite{Brown} offered several forgery attacks on our scheme. Brown and Monico also offered easy patches against both Panny's attacks but mentioned that they had not found any way to prevent one forgery attack of their own.

In this updated version of our original preprint \cite{original}, we take into account suggestions of Brown and Monico (both from \cite{Brown} and from informal communication) to thwart all these attacks, although to avoid the attack in Section 4.6 of \cite{Brown}, we had to modify our scheme in a more substantial way, see Section
\ref{alternative}.

Finally, we note that in \cite{Fiat-Shamir}, the authors offered an interactive version, in the spirit of Fiat-Shamir \cite{Fiat}, of our protocol in \cite{original}. This tool typically adds an extra level of security to any ZKP protocol. However, we prefer to stick to a non-interactive version here.

\section{Preliminaries}
\label{Preliminaries}

We start by giving some necessary information on tropical algebras
here; for more details, we refer the reader to the monograph
\cite{Butko}.

Consider a tropical semiring ${\bf A}$, also known as the min-plus algebra
due to the following definition. This semiring is defined as a linearly ordered set (e.g., a subset of reals) that contains 0 and  is closed under addition, with
two operations as follows:
\medskip

\noindent $x \oplus y = \min(x, y)$

\medskip

\noindent $x \otimes y = x+y$.
\medskip

It is straightforward to see that these operations satisfy the
following properties:
\medskip

\noindent {\it associativity}: \\
$x \oplus (y  \oplus z) = (x \oplus y)  \oplus z$\\
$x \otimes (y  \otimes z) = (x \otimes y)  \otimes z$.

\medskip

\noindent {\it commutativity}:\\
$x \oplus y = y \oplus x$\\
$x \otimes y = y \otimes x$.
\medskip

\noindent {\it distributivity}:\\
$(x \oplus y)\otimes z = (x \otimes z)  \oplus (y\otimes z)$.

\medskip

There are some ``counterintuitive" properties as well:

\noindent $x \oplus x = x$
\medskip

\noindent $x \otimes 0 = x$
\medskip

\noindent $x  \oplus 0$ could be either 0 or $x$.

\medskip

There is also a special ``$\epsilon$-element"  $\epsilon = \infty$
such that, for any $x \in S$,
\medskip

\noindent $\epsilon \oplus x = x$
\medskip

\noindent $\epsilon \otimes x = \epsilon$.

\subsection{Tropical polynomials}\label{polynomials}

A  (tropical)  monomial in $S$ looks like a usual linear function,
and a tropical polynomial is the minimum of a finite number of such
functions, and therefore a concave, piecewise linear function. The
rules for the order in which tropical operations are performed are
the same as in the classical case, see the example below. Still, we often use parenthesis to make a tropical polynomial easier to read.

\begin{example}
Here is an example of a tropical monomial: $x \otimes x \otimes y
\otimes z \otimes z$. The  (tropical)  degree of this monomial is 5.
We note that sometimes, people use the alternative notation
$x^{\otimes 2}$ for $x \otimes x$, etc.

An example of a   tropical polynomial is: $p(x, y, z) = 5\otimes x
\otimes y  \otimes z  \oplus x \otimes x \oplus 2\otimes z \oplus 17
=  (5\otimes x \otimes y \otimes z) \oplus (x \otimes x) \oplus
(2\otimes z) \oplus 17.$ This polynomial has  (tropical) degree 3,
by the highest degree of its monomials.
\end{example}

We note that, just as in the classical case, a tropical polynomial
is canonically represented by an ordered set of tropical monomials
(together with finite coefficients), where the order that we use
here is deglex.

We also note that some tropical polynomials may look ``weird":

\begin{example}\label{zero}
Consider the polynomial $p(x)= (0 \otimes x) \oplus (0 \otimes x \otimes x)$. All coefficients in this polynomial are 0, and yet it is not the same as the polynomial $q(x)=0$.

Indeed, $q(x) \otimes r(x) = r(x)$ for any polynomial $r(x)$. On the other hand, if, say, $r(x)= 2 \otimes x$, then  $p(x) \otimes r(x) = (2 \otimes x \otimes x) \oplus (2 \otimes x \otimes x \otimes x) \ne r(x)$.

\end{example}

In the following example, we show in detail how two tropical polynomials are multiplied and how similar terms are collected.

\begin{example}\label{multiply}
Let $p(x)= (2 \otimes x) \oplus (3 \otimes x \otimes x)$ and  $q(x)= 5 \oplus (1 \otimes x)$. Then $p(x) \otimes q(x)= [(2 \otimes x)\otimes 5] \oplus [(2 \otimes x)\otimes (1 \otimes x)] \oplus [(3 \otimes x \otimes x) \otimes 5] \oplus [(3 \otimes x \otimes x) \otimes (1 \otimes x)] = (7 \otimes x) \oplus (3 \otimes x \otimes x) \oplus (8 \otimes x \otimes x) \oplus (4 \otimes x \otimes x \otimes x) = (7 \otimes x) \oplus (3 \otimes x \otimes x) \oplus (4 \otimes x \otimes x \otimes x).$

\end{example}

In this paper, our focus is on one-variable tropical polynomials, although one can use multivariate tropical polynomials instead.

\section{Digital signature scheme description} \label{scheme}

Let ${\bf T}$ be the tropical algebra of one-variable polynomials over ${\bf Z}$, the ring of integers.

\noindent The signature scheme is as follows.

\medskip

\noindent {\bf Private:}  two polynomials $X, Y \in {\bf T}$ whose degrees sum up to $2d$, with all coefficients in the range $[0, r]$, where $d$ and $r$ are  parameters of the scheme.
\medskip

\noindent {\bf Public:}

\noindent -- polynomial $M = X \otimes Y$

\noindent -- a hash function $H$ (e.g., SHA-512) and a (deterministic) procedure for converting values of $H$ to one-variable polynomials from the tropical algebra ${\bf T}$ (see Section \ref{vector}).

\medskip

\noindent {\bf Signing} a message $m$:
\medskip

\begin{enumerate}

\item[S1.]  Apply a hash function $H$ to $m$. Convert $H(m)$ to a polynomial
$P$ of degree $d$ from the algebra ${\bf T}$ using a deterministic public procedure.
\medskip

\item[S2.] Select two random private polynomials $U, V \in {\bf T}$ such that
$deg(U)=deg(Y),\\
deg(V)=deg(X)$, with all coefficients of $U$ and $V$ in the range $[0, r]$. Denote $N = U \otimes V$.
\medskip

\item[S3.] The signature is the 4-tuple of polynomials
$(P, ~P \otimes X \otimes U, ~P \otimes Y \otimes V, ~N)$.

\end{enumerate}

\vskip 0.2 cm

\noindent {\bf Verification:}

\medskip

\begin{enumerate}

\item[V1.]  The verifier computes the hash $H(m)$ and converts $H(m)$ to a
polynomial $P$ of degree $d$ from the algebra ${\bf T}$ using a deterministic public procedure. This is done to verify that $P$ is the correct hash of the message.
\medskip

\item[V2.]  The verifier checks that the degrees of the polynomials $P \otimes X
\otimes U$ and $P \otimes Y \otimes V$ (the second and third polynomials in the signature) are both equal to $3d$, and the degree of the polynomial $N$ is equal to $2d$. If not, then the signature is not accepted.
\medskip

\item[V3.]  The verifier checks that neither $P \otimes X \otimes U$ nor $P
\otimes Y \otimes V$ is a constant multiple (in the tropical sense) of $P \otimes M$ or $P \otimes N$. If it is, then the signature is not accepted.
\medskip

\item[V4.]  The verifier checks that all coefficients in the polynomials
$P \otimes X \otimes U$ and $P \otimes Y \otimes V$ are in the range $[0, 3r]$, and all coefficients in the polynomial $N$ are in the range $[0, 2r]$.
If not, then the signature is not accepted.

\medskip

\item[V5.]  The verifier computes $W=(P \otimes X \otimes U) \otimes (P \otimes Y \otimes V)$.
The signature is accepted if and only if $W=P \otimes P \otimes M \otimes N$.

\medskip

\end{enumerate}

\vskip 0.2 cm

\noindent {\bf Correctness} is obvious since $W=(P \otimes X \otimes U) \otimes (P \otimes Y \otimes V) = P \otimes P \otimes (X \otimes Y) \otimes (U \otimes V) = P \otimes P \otimes M \otimes N$.

\begin{remark}
Step V2 in the verification algorithm is needed to prevent trivial forgery, e.g. signing by the triple of polynomials  $(P \otimes M, ~P \otimes N, ~N)$, in which case $(P \otimes M)\otimes (P \otimes N) =P \otimes P \otimes M \otimes N$.
\end{remark}

\begin{remark}
Here is how one can check whether or not one given tropical polynomial, call it $R(x)$, is a constant multiple (in the tropical sense) of another given tropical polynomial (of the same degree), call it $S(x)$.

Let $r_i \in \Z$ denote the coefficient at the monomial $x^{\otimes i}$ in $R(x)$, and $s_i \in \Z$ denote the coefficient at the monomial $x^{\otimes i}$ in $S(x)$. If $R(x) = c \otimes S(x)$ for some $c \in \Z$, then $r_i = s_i +c$ for every $i$. Here ``+" means the ``classical" addition in $\Z$.

Therefore, to check if $R(x)$ is a constant multiple of $S(x)$, one checks if $(r_i - s_i)$ is the same integer for every $i$.

\end{remark}

\section{Key generation and suggested parameters}\label{generation}

The suggested value of $d$ is 150.

The degree of the polynomial $X$ is selected uniformly at random from integers in the interval $[\frac{3}{4} d, \frac{5}{4} d]$. The degrees of other private polynomials are then determined from the conditions $deg(X)+deg(Y)=2d$, $deg(V)=deg(X)$, $deg(U)=deg(Y)$.

All coefficients of monomials in the polynomials $X, Y, U, V$ are selected uniformly at random from integers in the range $[0, r]$, where $r=127$. We emphasize that, in contrast with the ``classical" case, if the coefficient at a monomial is 0, this does not mean that this monomial is ``absent" from the polynomial.

\subsection{Safe keys}\label{safe}

Similar to the situation with the RSA modulus $n=pq$ where the private primes $p$ and $q$ should be ``safe primes" (i.e., $p-1$ and $q-1$ should not have small divisors other  than 2), in our situation the private polynomials $X, Y, U, V$ should not have any non-constant divisors (in the tropical sense). Otherwise, the forgery attack from \cite{Panny} may apply.

It is not immediately clear how to efficiently generate irreducible tropical polynomials of a given degree. One simple way to do this with high probability is
zeroing the first and last coefficients of $X$ and $Y$. There is an argument in  \cite{Kim} suggesting that a generic polynomial whose first and last coefficients are 0 is irreducible.

However, in contrast with the classical situation, even if $X$ and $Y$ are irreducible tropical polynomials, this does not necessarily imply that the only factors of $M=X \otimes Y$ are $X$ and $Y$, although with high probability $M$ will not have any factors of low degree if the degrees of both $X$ and $Y$ are high. This will make a brute force factorization of $M$ computationally hard.

In any case, questions related to factoring one-variable tropical polynomials need to be explored more to provide for a reliable way of generating safe keys for our scheme.

\subsection{Converting $H(m)$ to a tropical polynomial over $\Z$}\label{vector}

We suggest using a hash functions from the SHA-3 family, specifically SHA3-512. We assume the security properties of SHA3-512, including collision resistance and preimage resistance. We also  assume that there is a standard way to convert $H(m)$ to a bit string of length 512. Then a bit string can be converted to a tropical polynomial $P=P(x)$ over $\Z$ using the following {\it ad hoc} procedure.

Let $B=H(m)$ be a bit string of length 512. We will convert $B$ to a one-variable tropical polynomial $P$ of degree $d=150$ over $\Z$. We therefore have to select 151 coefficients for monomials in $P$, and we want to have these coefficients in the range [0, 127]. With 7 bits for each  coefficient, we need $151 \cdot 7 =1057$ bits in total.

\medskip

\noindent {\bf (1)} Concatenate 3 copies of the bit string $B$ to get a bit string of length 1536.
\medskip

\noindent {\bf (2)} Going left to right, convert 7-bit block $\#j$ to an integer and use it as the coefficient at the monomial $x^{\otimes j}$.
\medskip

\noindent {\bf (3)} After we use $7 \cdot 151 = 1057$ bits, all monomials in the polynomial $P=P(x)$ will get a coefficient.

\subsection{Multiplying two tropical polynomials}\label{Multiplying}
Let $R(x)$ and $S(x)$ be two one-variable tropical polynomials of degree $d$ and $g$, respectively.
We want to compute $R(x) \otimes S(x)$.

Note that a one-variable tropical monomial, together with a coefficient, can be represented by a pair of integers $(k, l)$, where $k$ is the coefficient and $l$ is the degree. Our goal is therefore to compute the coefficient at every monomial of degree from 0 to $d+g$ in the product
$R(x) \otimes S(x)$.

Suppose we want to compute the coefficient at the monomial of degree $m, ~0 \le m \le d+g$.
Then we go over all coefficients $r_i$ at the monomials of degrees $i \le m$ in the polynomial $R(x)$ and add (in the ``classical" sense) $r_i$ to $s_j$, where $s_j$ is the coefficient at the monomial of degree $j=m-i$ in the polynomial $S(x)$.

Having computed all such sums $r_i+s_j$, we find the minimum among them, and this is the coefficient at the monomial of degree $m$ in the polynomial $R(x) \otimes S(x)$.

\section{What is the hard problem here?}\label{problem}

The (computationally) hard problem that we employ in our construction is factoring one-variable tropical polynomials. This problem is known to be NP-hard, see \cite{Kim}.

{\it Since recovering the private tropical polynomials $X$ and $Y$ from the public polynomial $M=X \otimes Y$ is exactly the factoring problem, we see that inverting our candidate one-way function
$f(X, Y)=X \otimes Y$ is NP-hard.}

However, the private tropical polynomials $X$ and $Y$ are involved also in the signature. For example, from the polynomial $W=P \otimes X \otimes U$ the adversary can recover $X \otimes U$ because the polynomial $P$ is public. The polynomial $U$ is private, so it looks like the adversary is still facing the factoring problem. However, the adversary now knows two products involving the polynomial $X$, namely $X \otimes Y$ and $X \otimes U$. Therefore, we have a somewhat different problem here: finding a common divisor of two given polynomials.

This problem is easy for ``classical" one-variable polynomials over $\Z$. In particular, any classical one-variable polynomial over $\Z$ has a unique factorization (up to constant multiples) as a product of irreducible polynomials. In contrast, a one-variable tropical polynomial can have an exponential number of incomparable factorizations \cite{Kim}. Furthermore, it was shown in \cite{Kim} that two one-variable tropical polynomials may not have a unique g.c.d.  All this makes it appear likely that the problem of finding the g.c.d. of two (or more) given one-variable tropical polynomials is computationally hard. No polynomial-time algorithm for solving this problem is known. More about this in Section \ref{attacks}.

\section{Possible attacks}\label{attacks}

The most straightforward attack is trying to factor the tropical polynomial $M = X \otimes Y$ as a product of two tropical polynomials $X$ and $Y$. As we have pointed out before, this problem is known to be NP-hard \cite{Kim}. In our situation, there is an additional restriction on the degrees of $X$ and $Y$, to pass Step V2 of the verification procedure.

If one reduces the equation $M = X \otimes Y$ to a system of equations in the coefficients of $X$ and $Y$, then one gets a system of $2d+1$ quadratic (tropical) equations in $2(d+1)$ unknowns. With $d$ large enough, such a system is unapproachable; in fact, solving a system of quadratic tropical  equations is known to be NP-hard \cite{TT}. The size of the key space for $X$ and $Y$ with suggested parameters is $128^{300} = 2^{2100}$, so the brute force search is infeasible.

It is unclear whether accumulating (from different signatures) many tropical polynomials of the form  $M_i= X \otimes U_i$, with different (still unknown) polynomials $U_i$ can help recover $X$.
With each new $M_i$, the attacker gets (on average) $d+1$ new unknowns (these are coefficients of $U_i$) and $2d+1$ new equations. There is a well-known trick of reducing a system of quadratic equations to a system of linear equations by replacing each product of two unknowns by a new unknown. However, the number of pairs of unknowns increases roughly by $d^2$ with each new $U_i$. Therefore, a system of linear
equations like that will be grossly underdetermined, resulting in a huge number of solutions for the new unknowns, thus making solving the original system (in the old unknowns) hard, especially given the restrictions on the old unknowns tacitly imposed by Step V4 of the verification procedure.

As we have mentioned in the Introduction, Panny \cite{Panny} and Brown and Monico \cite{Brown} offered several forgery attacks on the original version \cite{original} of  our scheme. The only serious attack of those is the factorization attack from Section 4.6 of \cite{Brown}, which prompted us to make some changes not only in the key generation procedure but also in the scheme itself, see our Section \ref{alternative}.

\section{Performance and signature size} \label{performance}

For our computer simulations, we used Apple MacBook Pro, M1 CPU (8 Cores), 16 GB RAM computer.
Python code implementing the original version of the scheme \cite{original} is available, see \cite{Python}.

We note that a one-variable tropical monomial, together with a coefficient, can be represented by a pair of integers $(k, l)$, where $k$ is the coefficient and $l$ is the degree of the monomial. Then a one-variable tropical polynomial of degree, say, 150 is represented by 151 such pairs of integers, by the number of monomials. If $k$ is selected uniformly at random from integers in the range $[0, 127]$, then the size of such a representation is about 2000 bits on average. Indeed, 151 coefficient of the average size of 6 bits give about 900 bits. Then, the degrees of the monomials are integers from 0 to 150. These take up $(\sum_{k=1}^{7} k \cdot (2^k - 2^{k-1} +1)) + 8 \cdot (150-127) \approx 1000$  bits. Thus, it takes about 2000 bits on average to represent a single tropical polynomial with suggested parameters.

Since a private key is comprised of two such polynomials, this means that the size of the (long-term) private key in our scheme is about 4000 bits (or 500 bytes) on average.

The public key is a polynomial of degree 300. Coefficients in this polynomial are in the range $[0, 254]$. Using the same argument as in the previous paragraph, we estimate the size of such polynomial to be about 4500 bits (or 562  bytes) on average.

The signature is a 4-tuple of polynomials, one of them has degree 150, two of them have degree 450, and one has degree 300. Therefore, the signature size is about 16,000 bits (or 2000 bytes) on average.

In the table below, we have summarized performance data for several parameter sets, in the case where all private polynomials $X, Y, U, V$ have the same degree. Most columns are self-explanatory; the last two columns show memory usage during verification and during the whole process of signing and verification.

\bigskip

\hskip -0.7cm
\begin{tabular}{|p{1.5cm}|p{1.5cm}|p{1.7cm}|p{1.5cm}|p{1.5cm}|p{1.5cm}|p{1.3cm}|p{1.3cm}|}
 \hline
 \multicolumn{8}{|c|}{Performance metrics for various parameter values} \\
 \hline

   degree of private polynomials & range for coefficients in private polynomials & verification time (sec) & signature size (Kbytes) & public key size (Kbytes)& private key size (Kbytes) & memory usage, verification (Mbytes) & memory usage, whole process (Mbytes) \\
 \hline
    100 & [0,127] & <0.1&  1.3 & 0.37 & 0.33 & 0.4 & 0.4 \\
  \hline

    150 & [0,127] & 0.15 & 2 & 0.56 & 0.5 & 0.37 & 0.5 \\
 \hline
   200  & [0,127] & 0.25 &  2.6 & 0.74 & 0.67 & 0.47 & 0.6\\

 \hline

\end{tabular}

\section{Alternative signature scheme}\label{alternative}

To completely avoid the division attack in Section 4.6 of \cite{Brown}, we offer here a similar but different  signature scheme where tropical addition plays a more prominent role.

The private and public keys are the same as in the scheme in our Section \ref{scheme}.
The only difference is that here the hash $H(m)$ is converted to a tropical polynomial of degree $2d$, not $d$.
\medskip

\noindent {\bf Signing} a message $m$:
\medskip

\begin{enumerate}

\item[$S'1$.]  Apply a hash function $H$ to $m$. Convert $H(m)$ to a polynomial
$P$ of degree $d$ from the tropical algebra ${\bf T}$ using a deterministic public procedure.
\medskip

\item[$S'2$.] Select two random private polynomials $U, V \in {\bf T}$ such that
$deg(U)=deg(Y),\\
deg(V)=deg(X)$, with all coefficients of $U$ and $V$ in the range $[0, r]$. Denote $N = U \otimes V$.
\medskip

\item[$S'3$.] Select a random public polynomial $E$ of degree $3d$, with all coefficients in the range $[0, 3r]$.
\medskip

\item[$S'4$.] The signature is the following 6-tuple of polynomials:\\
$(P, ~P \oplus (X \otimes U), ~P \oplus (Y \otimes V), ~P \otimes [(X \otimes U) \oplus (Y \otimes V)] \oplus E,  ~N, ~E)$.

\end{enumerate}

\vskip 0.2 cm

\noindent {\bf Verification:}

\medskip

\begin{enumerate}

\item[$V'1$.]  The verifier computes the hash $H(m)$ and converts $H(m)$ to a
polynomial $P$ of degree $2d$ from the algebra ${\bf T}$ using a deterministic public procedure. This is done to verify that $P$ is the correct hash of the message.
\medskip

\item[$V'2$.]  The verifier checks that the degrees of the polynomials $P \oplus (X
\otimes U)$ and $P \oplus (Y \otimes V)$ (the second and third polynomials in the signature) are both equal to $2d$, the degree of the polynomial $N$ is equal to $2d$ as well, and the degrees of the remaining two polynomials are  equal to $3d$. If not, then the signature is not accepted.
\medskip

\item[$V'3$.]  The verifier checks that all coefficients in the polynomials
$P \oplus (X \otimes U)$ and $P \oplus (Y \otimes V)$ are in the range $[0, 2r]$,  all coefficients in the polynomial $N$ are in the range $[0, 2r]$ as well, and all coefficients in the remaining two polynomials are in the range $[0, 3r]$.
If not, then the signature is not accepted.
\medskip

\item[$V'4$.]  The verifier checks that neither $P \oplus (X \otimes U)$ nor $P
\oplus (Y \otimes V)$ is a constant multiple (in the tropical sense) of $P \oplus M$ or $P \oplus N$. If it is, then the signature is not accepted.
\medskip

\item[$V'5$.] Denote $R= P \otimes [(X \otimes U) \oplus (Y \otimes V)]$. The verifier checks that\\
$P \otimes [(P \oplus (X \otimes U)) \oplus (P \oplus (Y \otimes V))] \oplus E = (P \otimes P) \oplus (R \oplus E)$.\\
If not, then the signature is not accepted.

\medskip

\item[$V'6$.]  The verifier computes $W=(P \oplus (X \otimes U)) \otimes (P \oplus (Y \otimes V)) = (P \otimes P) \oplus (P \otimes [(X \otimes U) \oplus (Y \otimes V)]) \oplus (X \otimes U \otimes Y \otimes V) $.
The signature is accepted if and only if $W \oplus E =(P \otimes P) \oplus (R  \oplus E) \oplus (M \otimes N)$.

\medskip

\end{enumerate}

\subsection{Key generation}\label{generation2}

Key generation here follows Section \ref{generation}, except that the hash $H(m)$ should now be converted to a tropical polynomial of degree $2d$, not $d$. A simple way to do the latter is just to tropically multiply a polynomial of degree $d$ constructed as in Section \ref{vector}, by itself.

\subsection{Brown-Monico attack}\label{Brown}

The attack in Section 4.6 of \cite{Brown} is based on (tropically) dividing a public polynomial by another public polynomial. The result of such division is not unique, but it recovers the correct ratio with non-negligible probability.

Introducing tropical addition in the signature is intended as a countermeasure to this attack. Recovering $B$ from  $(A \oplus B)$ and $A$ is highly non-unique, so the probability of correctly recovering, say, $(X \otimes U)$ or $(Y \otimes V)$ from the polynomials in the signature has a lesser chance of being non-negligible.

\subsection{Performance and signature size} \label{performance2}
Speed of computation is not really different here from what it is for the scheme in Section \ref{scheme}.

The signature size though is about 50\% larger, so with $d=150, r=127$  it is about 3 Kbytes.

\vskip 0.5cm

\noindent {\bf Acknowledgement.} We are grateful to Dan Brown and Chris Monico for pointing out a couple of weaknesses in the original version of our scheme and for discussions/suggestions on safe keys.

\end{document}